\documentclass[%
reprint,
 twocolumn,
 showpacs,preprintnumbers,
 aps,
pra,
 longbibliography
]{revtex4-1}

\usepackage{amsmath}
\usepackage{mathbbol}
\usepackage{bbold}
\usepackage{mathrsfs}
\usepackage{graphicx}
\usepackage{dcolumn}
\usepackage{bm}
\usepackage[usenames]{color}
\usepackage[normalem]{ulem}

\usepackage[per-mode=symbol]{siunitx}   
\usepackage[version=3]{mhchem}
 
\DeclareSIUnit\intensity{\watt\per\centi\meter\squared}
\DeclareSIUnit\fieldstrength{\volt\per\centi\meter}
\DeclareSIUnit\kfieldstrength{k\volt\per\centi\meter}
\DeclareSIUnit\energy{cm^{-1}}

\newcommand{\melement}[3]{\ensuremath{\left\langle #1 \left|#2\right|#3\right\rangle}}%
\newcommand{\ie}{i.\,e.}%
\newcommand{\ket}[1]{\left|#1\right\rangle}
\newcommand{\bra}[1]{\left\langle #1\right|}

\newcommand{\expected}[1]{\left\langle #1\right\rangle}
\newcommand{\aarphys}{\affiliation{Department of Physics and Astronomy, Aarhus University, 8000  Aarhus C, Denmark}}%
\newcommand{\uoft}{\affiliation{Chemical Physics Theory Group, Department of Chemistry, and Center for Quantum Information and Quantum Control, University of Toronto, Toronto, ON M5S 3H6, Canada}}

\begin{document}

\title{Attosecond photoionization dynamics in neon}

\author{Juan J.\ Omiste}\uoft
\author{Lars Bojer Madsen}\aarphys

\date{\today}
\begin{abstract} 
We study the role of electron-electron correlation in the ground-state of Ne, as well as in photoionization 
dynamics induced by an attosecond XUV pulse.  For a selection of central photon energies around 
100 eV, we find that  while the mean-field time-dependent Hartree-Fock method provides qualitatively correct results for the total ionization yield, the photoionization cross section, the
photoelectron momentum distribution as well as for the time-delay in photoionization,
electron-electron correlation is important for a quantitative description of these quantities.
\end{abstract}
\maketitle
\section{Introduction}
\label{sec:introduction}

Ever since the early days of atomic physics, the role of electron-electron correlation has been a key topic. 
Even though ground and singly-excited states can be described by mean-field theory, it was soon
recognized that a correlated basis is beneficial for convergence to an accurate ground-state energy~\cite{Hylleraas1928,Hylleraas1929}. 
For multiply excited states, correlation is crucial and theory based on single-configurations breaks down~\cite{Fano1961}.
In strong-field and attosecond physics, the theoretical description is typically explicitly time-dependent 
and processes often involve  continua wavepackets carrying the temporal information encoded  by ultrafast pump and probe pulses~\cite{Krausz2009}. 
The presence of multiple continua challenges theory tremendously.
Needless to say carrying out a mean-field time-dependent Hartree-Fock calculation is possible for physical systems so large that
fully correlated configuration-interaction calculations are not. One of the tasks for theory is therefore to establish which level of approximation
is sufficient for a qualitative correct description of a given observable. This is one of the questions that we address in this work.
To this end we need a theory where we can control the level of approximation, and we need to investigate a system which is possible to describe at such different levels. Thus, we describe the photoionization of the Ne ground-state~\cite{Hochstuhl2014, Majety2015a,PhysRevA.96.022507} mediated by an ultrashort XUV pulse using the time-dependent restricted-active-space self-consistent-field (TD-RASSCF) approach~\cite{Miyagi2013,Miyagi2014,Miyagi2014b,Miyagi2017}. This method permits the introduction of restrictions on the 
number of excitations in the active orbital space and 
is a generalization of the multiconfigurational time-dependent Hartree-Fock method~\cite{Caillat2005}.
 In contrast to many-body methods based on time-independent single-particle orbitals such as time-dependent configuration-interaction with singles (TD-CIS)~\cite{Pabst2012}, time-dependent restricted-active-space configuration-interaction (TD-RASCI)~\cite{Hochstuhl2012} or time-dependent general-active-space configuration-interaction (TD-GASCI)~\cite{Bauch2014}, the TD-RASSCF is based on time-dependent single-particle orbitals that are
 optimally updated in each time-step. The latter approach benefits not only from a reduction in the number of orbitals,  but also from an implicit description of multiple ionization events, and from its flexibility to identify the most important configurations for an accurate description of the system~\cite{Miyagi2013,Miyagi2014b,Miyagi2017,Omiste2017_be,Leveque2016a}. We will study the ground-state energy and the photoionization dynamics and their
 sensitivity to the active orbital space, i.e., to the RAS scheme.  Finally, 
we will study the importance of electron correlation on the time-delay in photoionization. 
Time-delay studies have attracted much attention because of available experimental data~\cite{Schultze2010,Klunder2011} and 
the development of powerful theoretical and computational methods~\cite{Pazourek2015}. 
Time-delays have been evaluated in different scenarios, including in  photoionization from first principles~\cite{Su2013a}, as a function of angle of the ejected electron~\cite{PhysRevA.94.063409,PhysRevA.96.013408,Banerjee2017} or in strong-field ionization~\cite{Camus2017}. In particular, the experimental measurement of the time-delay in Ne between the photoemission of electrons from the $2s$ and $2p$ shells~\cite{Schultze2010} has provoked  a lot of interest due to a disagreement in the magnitude of the measured and calculated time-delays, see, e.g., Refs.~\cite{Kheifets2010,Baggesen2010,Komninos2011,Moore2011,Ivanov2011,Nagele2012,Su2014,Wei2016,Feist2014a,Pazourek2016}.
Note that very recently, analysis of  interferometric measurements~\cite{Isinger2017} suggests that a shake-up process,
not resolved and accounted for in the streaking experiment of Ref.~\cite{Schultze2010}, could affect the experimental result and 
possibly bring 
the experimental time-delay in agreement with the many-body calculations~\cite{Feist2014a}.
In this work, we propagate the photoelectron wavepacket to directly measure the time of emission from each electronic shell for a set of active spaces to probe the role of the electronic correlation~\cite{Omiste2017_be}.
 
The paper is organized as follows. In Sec.~II, we summarize the computational approach. 
In Sec.~\ref{sec:results}, we present our results. First, we study the ground-state for different RAS schemes. 
Next, we analyze the ionization dynamics induced by the laser. This analysis consists of the calculation of the ionization cross section and the description of the main features of the single photoionization, in particular,  the contribution of each ionization channel obtained by considering the photoelectron spectrum. Finally, we present computed time-delays between the electrons ejected from the $2s$ and $2p$ subshells and compare with available
experimental and  theoretical values. We conclude in Sec.~IV. Atomic units are used throughout unless indicated otherwise.

\section{Summary of the TD-RASSCF method}
\label{sec:td_ras_scf_method}

In this section, we summarize the TD-RASSCF method used to propagate the  many-electron wavefunction. We refer to previous works for details~\cite{Miyagi2013, Miyagi2014,Miyagi2014b,Omiste2017_be}. We propagate the dynamics of an $N_e$-electron atom in the laser field in the length gauge within the dipole approximation. The dynamics of this system is described by the Hamiltonian
\begin{eqnarray}
\nonumber
H&=&\sum_{j=1}^{N_e} \left(\cfrac{p_j^2}{2}-\cfrac{Z}{r_j}+\vec{E}(t)\cdot \vec{r}_j\right) +\sum_{j=1}^{N_e}\sum_{k>j}^{N_e}\cfrac{1}{|\vec{r}_j-\vec{r}_k|}=\\
  \label{eq:hamil}
&=&\sum_{j=1}^{N_e} h(\vec{r}_j,t) +\sum_{j=1}^{N_e}\sum_{k>j}^{N_e}\cfrac{1}{|\vec{r}_j-\vec{r}_k|},
\end{eqnarray}
where the first sum is over one-body operators and the second over two-body operators. The nuclear charge is denoted by $Z$ and the external electric field of the laser pulse is $\vec{E}(t)$.  To formulate and apply the TD-RASSCF theory, it is convenient to work in second quantization. We work in the spin-restricted framework, which implies that a given Slater determinant, $\ket{\Phi_{\bm{I}}(t)}$, is formed by $N_e/2$ spatial orbitals for each spin specie. In second quantization, the Hamiltonian reads 
\begin{equation}
  \label{eq:hamil_second}
H(t)= \sum_{pq} h^p_q(t)E^q_p +\frac{1}{2}\sum_{pqrs}v_{qs}^{pr}(t)E^{qs}_{pr},
\end{equation}
where we use the spin-free excitation operators ~\cite{Helgaker2000}
$E^q_p$ and $E^{qs}_{pr}$, defined as
$E^q_p=\sum\limits_{\sigma=\uparrow,\downarrow}b^\dagger_{p\sigma} b_{q\sigma},$ and $E^{qs}_{pr}=\sum\limits_{\sigma=\uparrow,\downarrow}\sum\limits_{\gamma=\uparrow,\downarrow}b^\dagger_{p\sigma}b^\dagger_{r\gamma} b_{s\sigma} b_{q\gamma}$
with $b^\dagger_{p\sigma}$ and $b_{p \sigma}$ the creation and annihilation operators of a single spin-orbital $\ket{\phi_p(t)}\otimes\ket{\sigma}$ and $\sigma$ denoting the spin degree of freedom. In Eq.~\eqref{eq:hamil_second}, the matrix elements are given by
\begin{eqnarray}
h_q^{p}(t)&=& \int\mathrm{d}\vec{r}\phi_p^*(\vec{r},t)h(\vec{r},t)\phi_q(\vec{r},t),\\
v_{qs}^{pr}(t)&=&  \int\int\mathrm{d}\vec{r}\mathrm{d}\vec{r'}\cfrac{\phi_p^*(\vec{r},t)\phi_r^*(\vec{r'},t)\phi_q(\vec{r},t)\phi_s(\vec{r'},t)}{|\vec{r}-\vec{r'}|}.
\end{eqnarray}
 The TD-RASSCF methodology is a generalization of MCTDHF~\cite{Beck2000a,Koch2006} in the sense that it includes the possibility to impose restrictions on the excitations in the active space~\cite{Miyagi2013,Miyagi2014b}, \ie, the many-body wavefunction reads
\begin{equation}
  \label{eq:wf_ansatz}
  \ket{\Psi(t)}=\sum_{\bm{I}\in\mathcal{V}} C_{\bm{I}}(t) \ket{\Phi_{\bm{I}}(t)},
\end{equation}
where the sum runs over the set of configurations $\mathcal{V}$, and not necessarily the full configuration space,
 and $C_{\bm{I}}(t)$ and $\ket{\Phi_{\bm{I}}(t)}$ are the amplitudes and Slater determinants of the configuration 
 specified by the index $\bm{I}$, which contains direct products of spin-up and spin-down strings, i.e., $\bm{I}=\bm{I_\uparrow}\otimes\bm{I_\downarrow}$, each of them including the indices of the spatial orbitals~\cite{Olsen1988,Klene2003}. Each Slater determinant is built from $M$ time-dependent spatial orbitals~$\{\phi_j(\vec{r},t)\}_{j=1}^M$, in the active orbital space $\mathcal{P}$. In the case of MCTDHF, $\mathcal{V}\equiv \mathcal{V}_\textup{FCI}$, that is, the full configuration space~\cite{Koch2006}. On the other hand, in the TD-RASSCF, the configurations are taken from  the restricted active space, $\mathcal{V}\equiv \mathcal{V}_\textup{RAS}$, which is defined as a subset of $\mathcal{V}_\textup{FCI}$ by imposing restrictions on the excitations in the active space. In this method, the active orbital space $\mathcal{P}$ is divided into 3 subspaces:~$\mathcal{P}_0,\,\mathcal{P}_1$ and $\mathcal{P}_2$. $\mathcal{P}_0$ constitutes the core, and its orbitals are fully occupied. All the different ways to form configurations by combination of orbitals in $\mathcal{P}_1$ are allowed. The orbitals in $\mathcal{P}_2$ are filled with restrictions by excitations from $\mathcal{P}_1$. The number of orbitals in $\mathcal{P}_0$, $\mathcal{P}_1$ and $\mathcal{P}_2$ are denoted by $M_0$, $M_1$ and $M_2$, and the total number of orbitals equals $M=M_0+M_1+M_2$.
The single-particle Hilbert space is completed by the ${\cal Q}$-space such that the unit operator can be resolved as 
$\mathbb{1} =P(t) + Q(t)$, with $P(t) = \sum_j | \phi_j (t) \rangle \langle \phi_j (t) |$ and $Q(t) = \sum_a | \phi_a(t)  \rangle \langle \phi_a(t) |$, with 	
$|\phi_j(t) \rangle$ belonging to $\mathcal{P}$-space and $|\phi_a(t) \rangle$ to 
${\mathcal Q}$-space.

In this work, we do not consider a core, \ie, we do not have a $\mathcal{P}_0$ subspace. We apply the TD-RASSCF-D method, \ie, include double (D) excitations from the active space partition $\mathcal{P}_1$ to $\mathcal{P}_2$. The TD-RASSCF-D method was shown to be numerically efficient and stable in the case of photoionization of Be~\cite{Omiste2017_be}. The equations of motion (EOM) read~\cite{Miyagi2013}
\begin{widetext}
  
  \begin{align}
    \label{eq:ci_dot}
    &i\dot C_{\bm{I}}(t)=\sum_{ij}\left[h^i_j(t)-i\eta_j^i(t)\right]\langle\Phi_{\bm{I}}(t)|E_i^j|\Psi(t)\rangle+\cfrac{1}{2}\sum_{ijkl}v_{jl}^{ik}(t)\langle\Phi_{\bm{I}}(t)|E_{ik}^{jl}|\Psi(t)\rangle,\\
    \label{eq:q_space}
    &i\sum_j Q(t) |{\dot\phi_j(t)}\rangle\rho_i^j(t)=Q(t)\left[\sum_jh(t)|{\phi_j(t)}\rangle\rho_i^j(t)+\sum_{jkl}W_l^k(t)|{\phi_j(t)}\rangle\rho_{ik}^{jl}(t)\right],\\
    \label{eq:p_space}
    &\sum_{k''l'}\left[h_{l'}^{k''}(t)-i\eta_{l'}^{k''}(t)\right]A_{k''i'}^{l'j''}(t)+\sum_{klm}\left[v_{kl}^{j''m}(t)\rho_{i'm}^{kl}(t)-v_{i'm}^{kl}(t)\rho_{kl}^{j''m}(t)\right]=0, 
  \end{align}
  with
  \begin{eqnarray}
    \eta_j^{i}(t)&=& \langle \phi_i(t)|\dot{\phi}_j(t)\rangle,\, Q(t)=\mathbb{1}-P(t)=\mathbb{1}-\sum_{j=1}^M \ket{\phi_j(t)}\bra{\phi_j(t)}\\
    W_l^k(\vec{r},t)&=&\int \phi_k^*(\vec{r'},t)\cfrac{1}{|\vec{r}-\vec{r'}|}\phi_l(\vec{r'},t)\mathrm{d}\vec{r'},\quad \rho_i^j(t)=\langle\Psi(t)|E_i^j|\Psi(t)\rangle,\\
    \label{eq:e_e_rho}
    \rho_{ik}^{jl}(t) &=&\langle\Psi(t)|E_{ik}^{jl}|\Psi(t)\rangle,\quad A_{ki}^{lj}(t)= \langle\Psi(t)|E_i^jE_k^l-E_k^lE_i^j|\Psi(t)\rangle,
    \label{eq:rho_a}
  \end{eqnarray}
\end{widetext}
where the orbitals denoted by single and double prime indexes belong to different partitions. The strategy to propagate the EOM is as follows. To propagate $| \Psi(t) \rangle$ we need  expressions for the time-derivative for the orbitals $|{\dot{\phi}_j(t)}\rangle$ and the time-derivative of the amplitudes $\dot{C}_{\bm{I}}(t)$. The time derivative of the orbital is split in $\mathcal{P}-$ and $\mathcal{Q}-$space contributions $|{\dot{\phi}_j(t)}\rangle=P(t)|{\dot{\phi}_j(t)}\rangle+Q(t)|{\dot{\phi}_j(t)}\rangle=\sum\limits_{i=1}\ket{\phi_i(t)}\langle{\phi}_i(t)|{\dot{\phi}_j(t)}\rangle+Q(t)|{\dot{\phi}_j(t)}\rangle=\sum\limits_{i=1}\eta_j^i\ket{\phi_i(t)}+Q(t)|{\dot{\phi}_j(t)}\rangle$. Equation~\eqref{eq:p_space} is used to determined the $\eta_j^i$'s. With these at hand the $P(t)|\dot{\phi}_j(t)\rangle$-part of the derivative  of the orbital is determined. Equation~\eqref{eq:q_space} is then used to find $Q(t)|\dot{\phi}_j(t)\rangle$ and finally $\dot{C}_{\bm{I}}(t)$ is determined from Eq.~\eqref{eq:ci_dot}.
The bottleneck of the propagation lies in the update at every time step of the two-body operator.
In order to speedup this update, we recently derived and described the coupled basis method~\cite{Omiste2017_be}. This method consists of coupling the angular part of the single-electron orbitals. It is easy to see that the angular momentum and magnetic quantum numbers of the coupled angular momenta are preserved by the two-body operator, $\left[1/|\vec{r}-\vec{r'}|,(\vec{\ell}+\vec{\ell'})^2\right]=\left[1/|\vec{r}-\vec{r'}|,\ell_z+\ell'_z\right]=0$, where $\vec{\ell}~(\vec{\ell'})$ and $\ell_z~(\ell'_z)$ are the one-electron angular momentum operators. This conservation property, together with the description of the radial part by a  finite-element discrete-variable-representation (FE-DVR), significantly reduces the number of operations in the evaluation~\cite{Omiste2017_be}. 

\section{Results}
\label{sec:results}

In this section, we describe the many-electron wavefunction of the ground-state of Ne and the photoionization dynamics including time-delay studies, induced by the interaction with an XUV linearly polarized attosecond pulse. In our simulations, we set the maximum angular momentum to $\ell_\text{max}=3$ and the maximum magnetic quantum number of each orbital to $m_\text{max}=2$. For the ground-state studies, the localization of the wavefunction near the nucleus allows us to confine the extend of the radial box to the interval $r\in [0,31)$, where we use 12 equidistant elements for $0\le r\le 6$ and complete the box with 10 equidistant elements up to $r=31$. The description of the photoionization process requires a larger box, which we build by adding 68 elements of 2.5~atomic units of length up to $r_\text{max}=201$. Each element contains 8 nodes. The results has been checked against convergence with respect to the parameters
of the primitive basis.

\subsection{Ground-State}
\label{sec:groundstate}

The TD-RASSCF method is developed to solve the TDSE for problems that are so large that a diagonalization of the Hamiltonian is impossible. As a consequence, the eigenstates of the system cannot in general be obtained straightforwardly. However, propagation in imaginary time of an initial guess function makes it possible to obtain the ground-state, since contributions of excited states are removed after long enough propagation time. The  nonlinearity of the EOM, Eqs.\eqref{eq:ci_dot}-\eqref{eq:p_space}, together with the 3D nature of the system demands an appropriate selection of the guess function to facilitate the convergence to the ground-state~\cite{Omiste2017_be}. Specifically, we choose the partitions $(M_1,M_2)=~(5,0),\,(5,1),\,(6,0),\,(5,4),\,(1,8),\,(9,0)$ and $(5,9)$, 
where the orbitals in the initial guess function are chosen as the hydrogenic orbitals $1s,2s,2p,3s,3p$ and $3d$ with nuclear charge $Z=10$. In addition, we impose that only the set of configurations $\left\{\ket{\Phi_{\bm{I}}(t)}\right\}$ with total magnetic quantum number $M_L=0$ contribute to $\ket{\Psi(t)}$ of Eq.~\eqref{eq:wf_ansatz}. This choice leads to stable numerical performance, because the initial set of orbitals and the restriction in $M_L$ ensures the correct symmetry of the ground-state. In addition, these two constrains imply that the magnetic quantum number of each orbital is conserved by the EOM, as shown in Appendix~\ref{sec:conservation_or_orbital_m} (see also Refs.~\cite{Kato2008,Haxton2012,Sato2016a}).
\begin{table}\centering
\caption{\label{tab:ground_state_energies} Ground-state energies of Ne for several RAS schemes. $M_1$ and $M_2$ denote the numbers of orbitals in the active orbital subspaces $\mathcal{P}_1$ and $\mathcal{P}_2$, respectively. When all the orbitals are in $\mathcal{P}_1$ ($M_1=M$ and $M_2=0$), the TD-RASSCF-D method corresponds to MCTDHF with $M_1$ orbitals. The $M_1=M=N_e/2$ case corresponds to TDHF.}
\begin{ruledtabular}
  \begin{tabular}{ccccc}
    Number of orbitals & $M_1$ & $M_2$ &Number of & ground-state\\
    $M=M_1+M_2$ &  &  & configurations & energy (a.u.)\\
    \hline
    5 & 5  & 0 & 1 & -128.548\\
    \hline
    6 & 5 & 1 & 26 & -128.561\\
      & 6 & 0 & 36 & -128.561\\
    \hline
    9 & 5 &4&521& -128.679\\
      & 1& 8& 8036& -128.682 \\
      & 9& 0&15876& -128.683 \\     
    \hline
    14 & 5 & 9&2746 & -128.765\\
  \end{tabular}
\end{ruledtabular}
\end{table}

We show the ground-state energies obtained by imaginary time propagation in Table~\ref{tab:ground_state_energies} for the considered RAS partitions.
The ground-state energy for the Hartree-Fock method $(5,0)$ coincides with previous calculations~\cite{Bunge1993b}. Note that for $M=6$, the TD-RASSCF-D with $(5,1)$ and $(6,0)$ are theoretically equivalent, because $M=N_e/2+1=6$. Therefore, we obtain the same ground-state energy and, as we will see in the next section, the photoionization dynamics induced by linearly polarized light is the same~\cite{Miyagi2013}. The strength of the TD-RASSCF method clearly manifests itself in the case of 9 orbitals. The energy difference between $(5,4)$ and $(9,0)$ is approximately $0.004$~a.u., whereas the number of configurations considered in the MCTDHF is 30 times larger than for the TD-RASSCF-D. The ground-state energy for $(5,9)$ is better than in the case of $(9,0)$ although the number of configurations is 6 times smaller. Let us remark that a smaller number of configurations does not necessarily mean a smaller numerical effort for a given number of orbitals, $M$, since the number of operations required to calculate the two-body operator scales as $\sim O(M^4)$~\cite{Omiste2017_be}. However, for a given number of orbitals, the MCTDHF calculation requires much more memory  for storing the amplitudes than the TD-RASSCF method. For instance, MCTDHF with 14 orbitals consists of 4008004 configurations, approximately 1459 times more than 
the $(M_1,M_2)=(5,9)$ case.
 
\subsection{Ionization and photoelectron spectrum}
\label{sec:photoelectron_spectrum}
\begin{figure}
  \includegraphics[width=.95\linewidth]{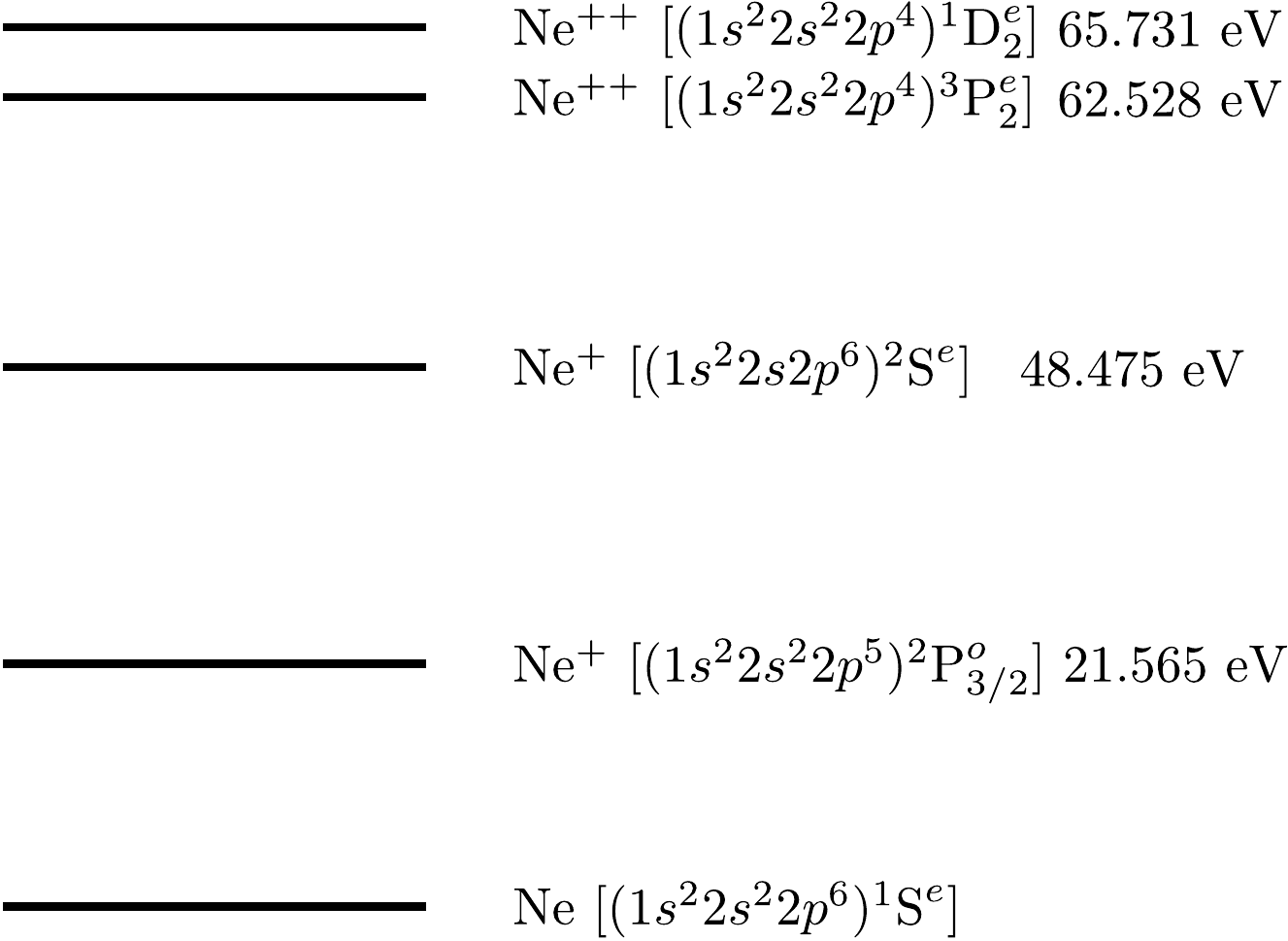}
  \caption{\label{fig:fig1} Ne ground-state and predominant Ne$^+$ and  Ne$^{++}$ channels for photons with $75~\text{eV}\le\omega\le 115$~eV. Note that we only show the lowest energy state in each fine-structure multiplet. The data are taken from Refs.~\cite{Saloman2004,Persson1971,Kramida2006}.}
\end{figure}

In this section, we investigate the ionization dynamics of Ne induced by an XUV laser pulse. We consider a laser pulse that is linearly polarized along the $Z$ axis of the laboratory frame and given by the vector potential $\vec{A}(t)=A_0\hat{z}\cos^2\left[\omega t/(2n_p)\right]\sin\omega t$, where $\omega$ and $n_p $ are the angular frequency and the number of cycles, and the duration is given by $T=2\pi n_p/\omega$. The pulse begins at $t=-T/2$ and ends at $T/2$. We set the intensity of the pulse to $10^{14}$~W/cm$^{2}$ and choose $n_p=10$ for $\omega$ in the range 75 to 115~eV. For this photon energy range, the predominant ionization channels are Ne${}^+(1s^22s2p^6)^2\text{S}^e$ and Ne${}^+(1s^22s^22p^5)^2\text{P}^o$~\cite{Saloman2004,Persson1971,Kramida2006}, see Fig.~\ref{fig:fig1}. The double ionization threshold is at 62.53~eV corresponding to the channel Ne${}^{++}(1s^22s^22p^4)^3\text{P}^e$.

\begin{figure}
  \includegraphics[width=.95\linewidth]{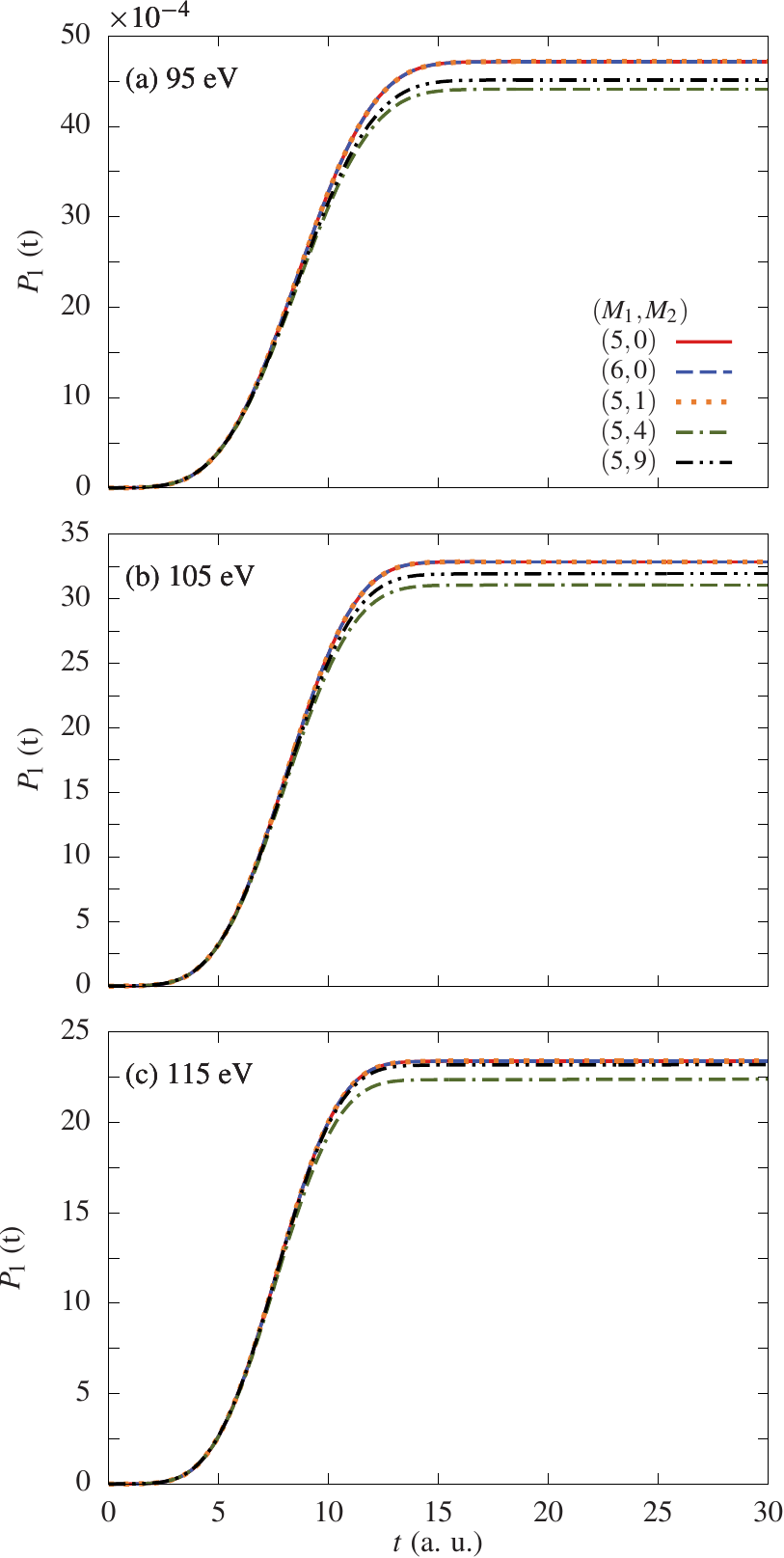}
  \caption{\label{fig:fig2} Total ionization yield as a function of time, $P_1(t)$, for a 10 cycle linearly polarized pulse with peak intensity $10^{14}$~W/cm$^2$ for (a) $\omega = 95$, (b) $105$ and (c) $115$~eV as a function of time for several RAS partitions. Note that the maximum of the pulse is at $t=0$.}
\end{figure}

We show the ionization yield as a function of time, $P_1(t)$ for $\omega=95,\,105$ and $115$~eV, in Fig.~\ref{fig:fig2}. The ionization yield is determined from the electron density in the outer region,  $P_1(t)=\sum_{ij}\rho_{i}^j\int_\Omega\mathrm{d}\Omega\int_{r_{out}}^\infty \phi_i(\vec{r},t)^*\phi_j(\vec{r},t)\mathrm{d} r$,  with $r_{out}=20$~a.u.
First, let us remark that the ionization yield calculated using TDHF, MCTDHF with 6 orbitals or the TD-RASSCF-D method for the RAS $(M_1,M_2)=(5,\,1)$ give the same result, showing that for linearly polarized lasers these approaches are equivalent as was the case in the previous section for the ground-state studies. Therefore, in the rest of this paper we only present the TDHF results to illustrate these three cases. 

\begin{figure}
  \includegraphics[width=.95\linewidth]{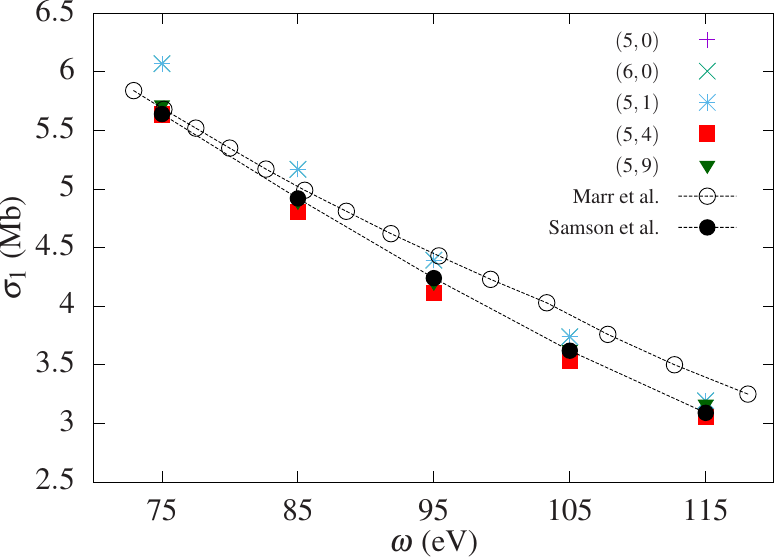}
  \caption{\label{fig:fig3} Total photoionization cross section calculated for a 10 cycles linearly pulse with peak intensity $10^{14}$~W/cm$^2$ as a function of the central frequency $\omega$ for several RAS partitions compared to the experimental data by Marr et al.~\cite{Marr1976} and Samson et al.~\cite{Samson2002}. The outcome for TDHF and RAS $(5,1)$ and $(6,0)$ are indistinguishable (see text). }
\end{figure}

In Fig.~\ref{fig:fig2}(a) we show the ionization yield for $\omega=95$~eV. At $t=0$, \ie, at the maximum of the laser field, the detected ionization is close to zero, and increases monotonically with time as the photoelectron wavepacket escapes from the inner region. For all the RAS partitions used, at approximately $t=15$~a.u., $P_1(t)$ reaches a plateau, which means that the electron wavefunction is beyond $r_{out}$. We see that the ionization yield obtained by the TDHF method is higher than the yields obtained with the other methods, which include correlation by populating $\mathcal{P}_2$. In numbers, for TDHF, the plateau-value for the ionization probability is $P_1\approx 47.18\times 10^{-4}$, and it reduces to $44.13\times 10^{-4}$ and $45.12\times 10^{-4}$ for $(M_1,M_2)=(5,4)$ and $(5,9)$, respectively. For $\omega=105$~eV [Fig.~2(b)], the pattern is the same, but $P_1(t)$ in the plateau region is smaller, $32.86\times 10^{-4}$ for TDHF, since the photon energy is further from the ionization energy than $\omega=95$~eV. 
We see in Fig.~\ref{fig:fig2}(c) that $P_1(t)$ for $\omega=115$~eV is much smaller for TD-RASSCF with $(M_1,M_2)=(5,4)$ than for TDHF, whereas the partition $(M_1,M_2)=(5,9)$ results in an 
ionization yield similar to that obtained with  TDHF.  Accordingly, the effect of  electron correlation cannot be 
clearly identified from the inspection of total ionization yields. A deeper understanding necessitates 
the study of a more differential quantity such as  the photoelectron spectrum (PES), as we will come back to below. First, however, 
we compare the TD-RASSCF calculation of the ionization with the experimental cross section~\cite{Marr1976,Samson2002}, $\sigma$, for $75~\text{eV}\le \omega\le 115$~eV in Fig.~\ref{fig:fig3}. We remark that the experimental data are very similar, although the data from Ref.~\cite{Marr1976} are systematically larger than those  from Ref.~\cite{Samson2002}. To extract the single ionization cross section, we use the following expression~\cite{Foumouo2006}
\begin{equation}
\label{eq:sigma_expression}
  \sigma_1 (\text{Mb})=1.032\times 10^{14}\omega^2 P_1/(n_p I_0),
\end{equation}
where  $I_0$ is the peak intensity of the laser pulse in W/cm${}^{2}$ and $P_1$ corresponds to $P_1(t)$ at a time when a constant value is reached. Equation \eqref{eq:sigma_expression} is valid for 
one-photon processes, and estimates the effective interaction time to be $T_\text{eff}={3\pi n_p}/{4\omega}$~\cite{Madsen2000,Foumouo2006}. In Fig.~\ref{fig:fig3}, we see that the TDHF method overestimates the cross section, although the difference with the experiment decreases with increasing photon energy. On the other hand, for $\omega=75$~eV, the results of the $(M_1,M_2)=(5,4)$ and $(5,9)$ calculations coincide with the experimental results of Ref.~\cite{Samson2002}. As we increase $\omega$, TD-RASSCF with $M_2=4$ underestimates the cross section up to $115$~eV, where the difference with the experimental value decreases. The numerical calculation using $M_2=9$ is in better agreement with the experimental result, which manifests the key role of the electron correlation in the photoionization process of Ne~\cite{Majety2015a}. The agreement is better than that obtained 
with other methods which cannot fully account for correlation effects~\cite{Hochstuhl2013,Hochstuhl2014}.  

\begin{figure}
  \includegraphics[width=.95\linewidth]{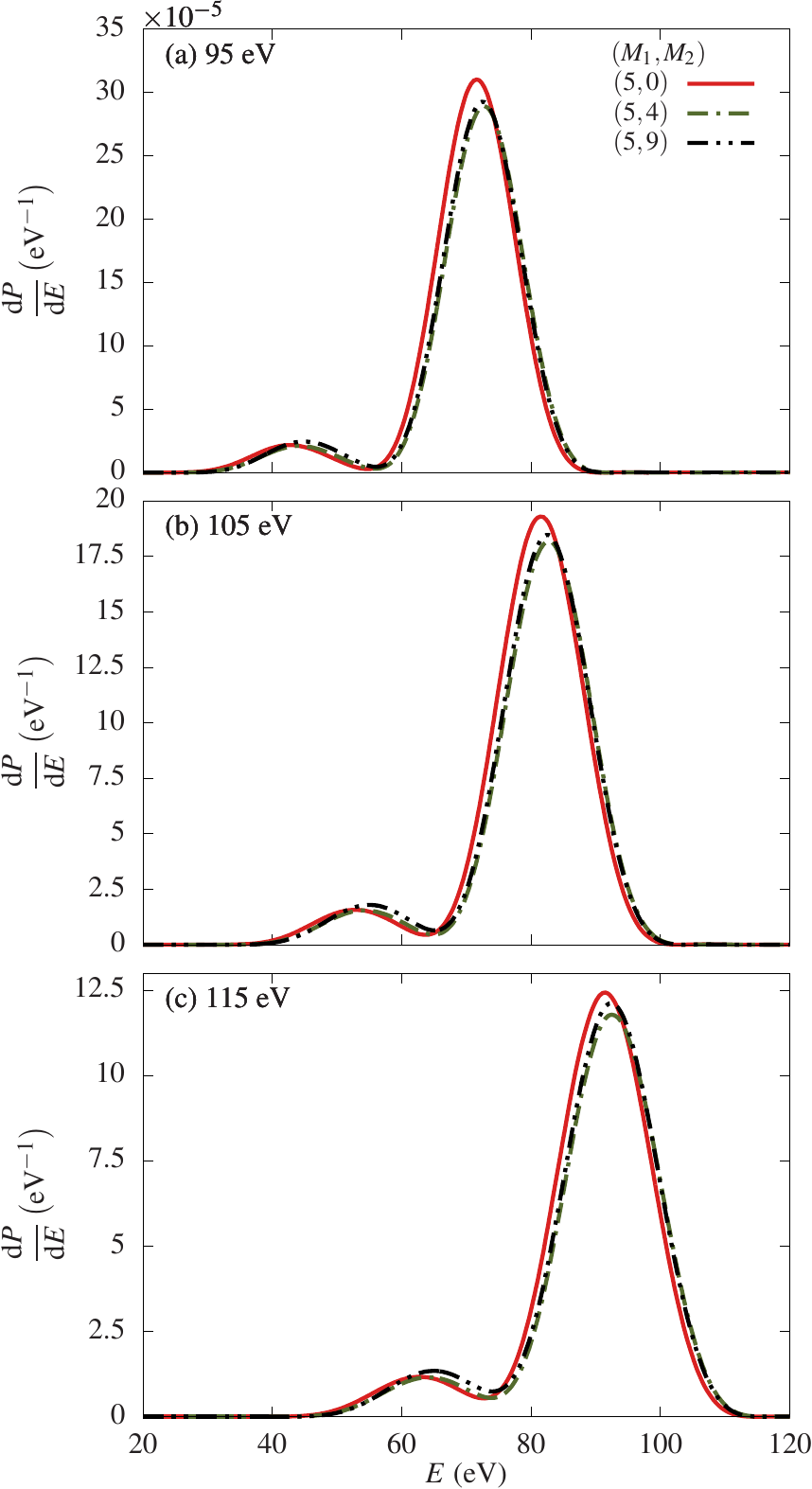}
  \caption{\label{fig:fig4} Photoelectron spectra as a function of the emitted electron energy for a 10 cycle linearly polarized pulse with peak intensity $10^{14}$~W/cm$^2$ for (a) $\omega = 95$, (b) $105$ and (c) $115$~eV as a function of time for several RAS partitions.}  
\end{figure}
Finally, we compare the PES for $\omega=95,\,105$ and $115$~eV as a function of the energy of the ejected electron in Fig.~\ref{fig:fig4}. We obtain the PES by considering the projection of the wavefunction for $r_{out}\ge 20$ on Coulomb waves~\cite{Madsen2007} using a window function to bypass boundary effects associated with the inner region and the end of the box. The procedure is explained in Ref.~\cite{Omiste2017_be}. At a given photon energy, the main peak corresponds to an electron ejected from the $p$ shell of the ground-state of Ne~$[(1s^22s^22p^6){}^1\text{S}^e]$, where the ion is left in the state Ne$^+~[(1s^22s^22p^5){}^2\text{P}^o]$ whose energy is $21.565$~eV~[Fig.~\ref{fig:fig1}]. The peak at lower energy corresponds to ionization into the channel Ne$^+~[(1s^22s2p^6){}^2\text{S}^e]$, at $48.475$~eV~[Fig.~\ref{fig:fig1}]. For all the photon energies considered in Fig.~\ref{fig:fig4}, the TDHF method overestimates the height of the main peak with respect to the TD-RASSCF using $(M_1,M_2)=(5,4)$ and $(5,9)$, and, moreover, the position of the peak is located at lower energies for the TDHF method. 
For $\omega=95$~eV, the dominant peak is quite similar for both TD-RASSCF schemes, but, as we increase the photon energy the maximum height of the peak corresponding to the $(5,9)$ partition becomes slightly larger than in the case of $(5,4)$ and the position of the peak shifts to higher energies. Specifically, the peaks are located at $42.47$ and $71.53$~eV for TDHF, at $44.087$ and $72.8$~eV for $(M_1,M_2)=(5,4)$ and  at $44.87$ and $72.40$~eV for $(M_1,M_2)=(5,9)$, compared to the experimental values of $46.525$ and $73.435$~eV~\cite{Saloman2004,Kramida2006}. Let us remark that as we increase the photon energy, since we set the number of cycles to $n_p=10$,  the distributions are wider due to the broadening of the spectral components of the laser pulse ($\Delta \omega\propto \omega/n_p$ with $\omega$ the central angular frequency). The tails of the distributions induced by this broadening does, however, not affect the position of the neighboring peaks for the frequencies used in this work. For instance, for $\omega=115$~eV and using the RAS partition $(M_1,M_2)=(5,9)$, the ionization threshold (extracted from the position of the peaks of the PES) are $22.67$ and $50.16$~eV for Ne$^+~[(1s^22s^22p^5){}^2\text{P}^o]$ and Ne$^+~[(1s^22s2p^6){}^2\text{S}^e]$, respectively, which agree with the thresholds extracted using the same RAS but $\omega=95$~eV.

\subsection{Time-delay}
\label{sec:time_delay}

In this section we calculate the time-delay between the ejection of electrons from the $2s$ and $2p$ subshells of Ne after the interaction with the pulse and compare with the experimental value obtained using the streaking technique~\cite{Schultze2010} and values from theory~\cite{Kheifets2010,Moore2011,Nagele2012,Su2014,Feist2014a,Pazourek2016,Wei2016}. Our strategy consists of extrapolating the streaking time-delay,~$\tau$, from the effective ionization time of the photoelectron ejected from a given subshell, $t_\text{Coul}(t)$, that we can extract from the dynamics of the electrons in the outer region. It reads as 
\begin{equation}
  \label{eq:time_delay_coul}
  t_\text{Coul}(t)=t-\cfrac{\expected{r(t)}}{k},
\end{equation}
with ${\expected{r(t)}}$ the expectation value of the position in the outer region at a given time $t$. Let us note that the apparent ionization time-delay $t_\text{Coul}(t)$ depends on time $t$ because in the presence of the Coulomb tail of the ion, the photoelectron cannot be described by a field-free wavepacket in the outer region. We can separate the dependence on $t$ using the relation~\cite{coulomb_phase_shift}
\begin{equation}
  \label{eq:time_delay_ews}
  t_\text{Coul}(t)=\tau_\text{EWS}+\Delta t_\text{Coul}(t)
\end{equation}
where  $\tau_\text{EWS}$ is the Eisenbud-Wigner-Smith (EWS) time-delay, which corresponds to the time required to escape the potential without the interaction with the Coulomb tail and $\Delta t_\text{Coul}(t)=\cfrac{Z}{k^{3}}\left[1-\ln(2k^2t)\right]$ is the distortion caused by the long range nature of the Coulomb potential. In Eq.~\eqref{eq:time_delay_ews}, $k$ is the linear momentum of the photoelectron and $Z=1$ is the charge of the remaining Ne$^+$ ion. Due to the short duration of the ionizing pulse, the photoelectron is described by a wavepacket in $k$, and hence $k$ attains several values. It is possible to describe the distribution over $k$, e.\,g., by $\expected{k}$ or $\expected{k^2}^{1/2}$, where $\langle\rangle$ denotes expectation value. In this work, we calculate $k$ by solving Eq.~\eqref{eq:time_delay_ews} for two different times. In the streaking experiments it is not $\tau_\text{EWS}$ that is measured directly, but rather the streaking time-delay, $\tau$, which may be written as
\begin{equation}
\label{eq:time_delay}
  \tau=\tau_\text{EWS}+\tau_\text{CLC},
\end{equation}
 where $\tau_\text{CLC}$ is the contribution due to the Coulomb-laser coupling~\cite{Pazourek2015}, and corresponds to the interaction with the IR field used in the streaking scheme. The quantity $\tau_\text{CLC}$ can be extrapolated accurately by~\cite{Feist2014a,Pazourek2015}
 \begin{equation}
\label{eq:time_delay_clc}
   \tau_\text{CLC}= \cfrac{Z}{k^{3}}\left[2-\ln\left(\cfrac{\pi k^2}{\omega_\text{IR}}\right)\right],
 \end{equation}
where $\omega_\text{IR}$ is the frequency of the IR pulse~\cite{Pazourek2015}. In our study, we do not include any streaking field, 
 but to compare with experiments $\tau_\text{CLC}$ has to be accounted for. 
 In this work, typical values for $\tau_\text{CLC}$ range from $\sim 9$~as for a photon energy of $\omega=85$~eV to $\sim 2.3$~as for 
 a photon energy of $\omega=125$~eV.

Next, we discuss the numerical method used to obtain the relative time-delay between the photoionization from $2s$ and $2p$ subshells, \ie, $\tau_{2p-2s}=\tau_{2p}-\tau_{2s}$. The photoionization channels involved are, in terms of dominant configurations, 
\begin{eqnarray}
\nonumber
\text{Ne}~[(1s^22s^22p^6){}^1\text{S}^e]&\rightarrow&\text{Ne}^+~[(1s^22s^22p^5){}^2\text{P}^o]+e^-(s,d)\\
  \label{eq:2p_ejected}
&&\\
  \nonumber
\text{Ne}~[(1s^22s^22p^6){}^1\text{S}^e]&\rightarrow&\text{Ne}^+~[(1s^22s2p^6){}^2\text{S}^e]+e^-(p),
\label{eq:2s_ejected}
\\
\end{eqnarray}
where the angular momentum $\ell$ of the emitted electron, $e^-$, is restricted to $s$ and $d$ when the electron is removed from the $2p$ subshell and to $p$ in the case of ionization of the $2s$ shell. We note that both channels in Eqs.~(\ref{eq:2p_ejected})-(\ref{eq:2s_ejected}) only involve the change of a single orbital, and they can hence both occur within the TDHF  description.

We can benefit from the difference in $\ell$ in the final continuum states to distinguish between the ionization channels by calculating $\expected{r(t)}$ along the parallel (all the three channels contribute) and perpendicular direction (only the $s$ and $d$ contribute) to the polarization of the laser. This technique was successfully applied in the case of Be, and relied on the fact that the contribution of the $s$ and $d$ photoelectrons is negligible with respect to that of the $p$ electron~\cite{Omiste2017_be}. 
\begin{figure}
  \includegraphics[width=\linewidth]{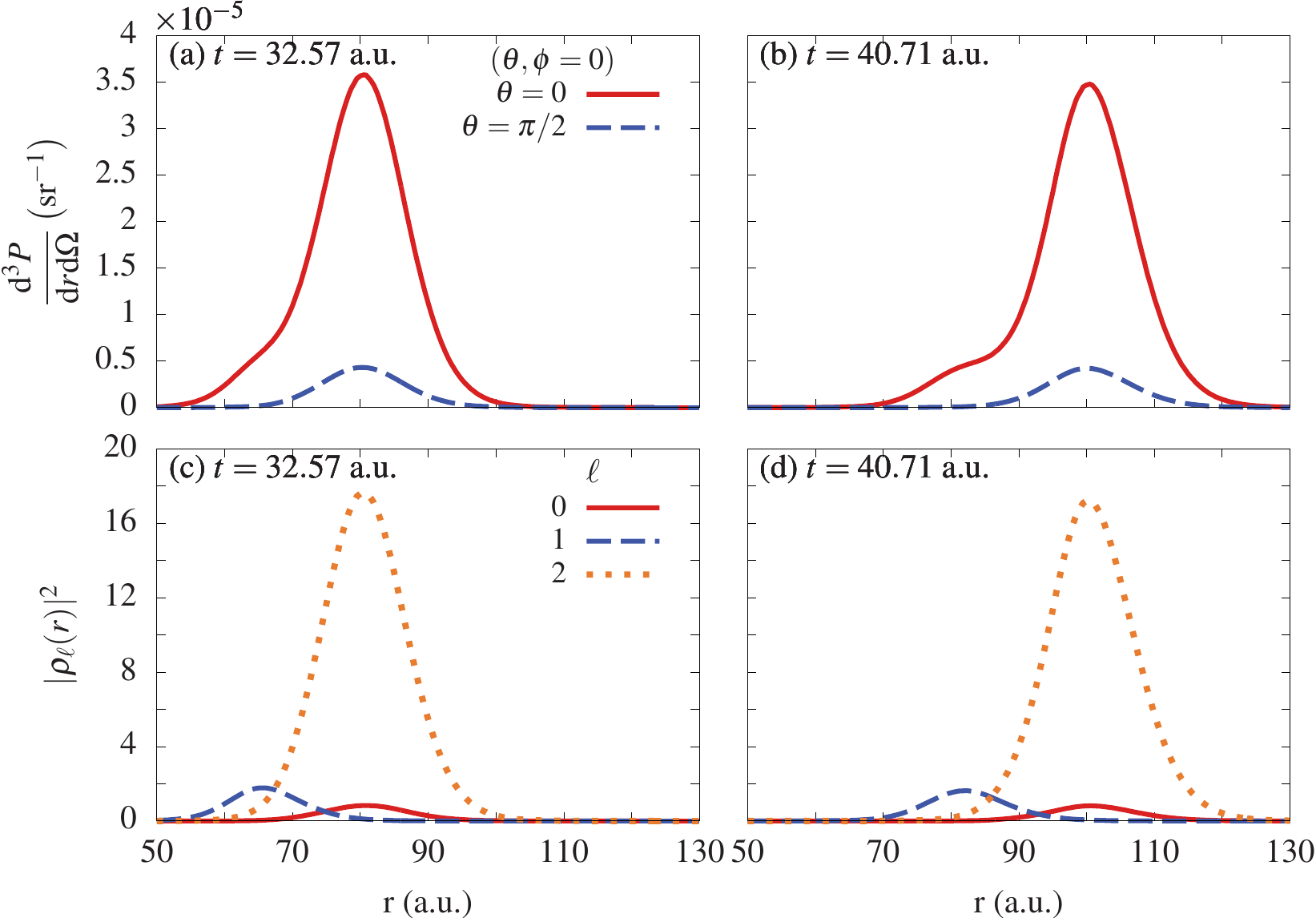}
  \caption{\label{fig:fig5} Triple differential density at different times after the peak of the pulse with $\omega=105$~eV (a) $t=32.57$~a.u. and (b) $t=40.71$~a.u. and the radial density of the components with a given $\ell$ at (c) $t=32.57$~a.u. and (d) $t=40.71$~a.u. for the photoelectron wavepacket for the TD-RASSCF method with $(M_1,M_2)=(5,4)$. }  
\end{figure} 
This is, however, not the case for Ne, since the cross section for photoionization from the $2p$ shell (\ie, $s$ and $d$ continuum electrons) is much larger than that from the $2s$ shell (\ie, $p$ continuum electrons~\cite{Hochstuhl2014}). We illustrate the implications of this difference in the outgoing wavepacket in Figs.~\ref{fig:fig5}(a)-(b), where we show the triple differential density (TDD) along the parallel and perpendicular directions for two different times after the peak of the XUV pulse with a central frequency of $\omega=105$~eV. The distribution along the parallel direction at $t=32.57$~a.u. is dominated by the emission of $s$ and $d$ photoelectrons, and it is peaked at approximately $r=80.5$~a.u. and presents a small shoulder around 60~a.u., which corresponds to the $p$ continuum electron contribution. At a later time, $t=40.71$~a.u., this shoulder becomes more pronounced because the $p$ photoelectron is slower than the $s$ and $d$ electron provoking the separation of the contributions, simply because of the differences in the ionization thresholds in Eqs.~\eqref{eq:2p_ejected} and \eqref{eq:2s_ejected}. 
Consequently, to distinguish both channels in this direction is not numerically efficient or even feasible in the present case. In particular, because it would be necessary to i) propagate for longer times and ii) employ a larger radial box to avoid boundary effects. Furthermore, even if we could meet these two demands, it may be not possible to determine the expected position of the photoelectron, due to the spreading of the density as a function of time. On the other hand, the TDD along the perpendicular direction corresponds to removing the $2p$ electron, therefore we could calculate directly $\expected{r(t)}$ for this channel. 

\begin{figure}
  \includegraphics[width=.95\linewidth]{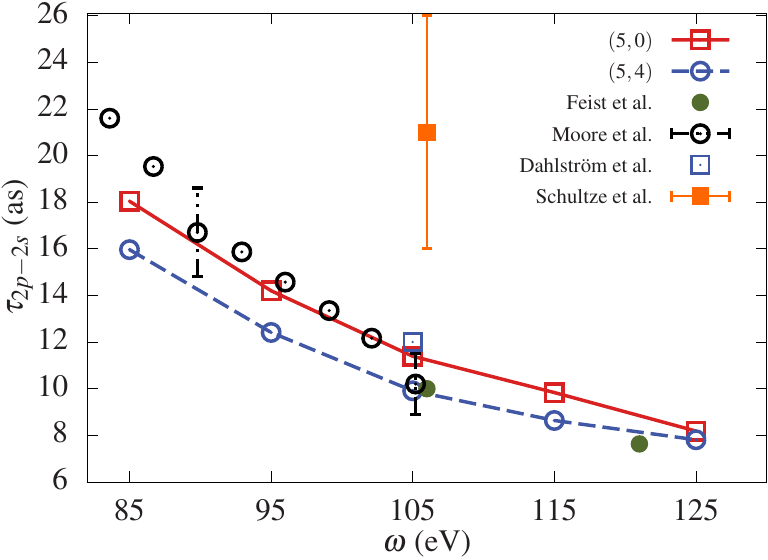}
  \caption{\label{fig:fig6} Relative time-delay of ionization in Ne, $\tau_{2p-2s}$, as a function of the central frequency of the XUV pulse for a 780~nm IR pulse for $(M_1,M_2)=(5,\,0)$ and $(5,\,4)$ together with the calculations by Feist et al.~\cite{Feist2014a}, Moore et al.~\cite{Moore2011} and Dahlstr\"om et al.~\cite{PhysRevA.86.061402} and the measurement by Schultze et al.~\cite{Schultze2010}.} 
\end{figure}

To overcome these difficulties imposed by determining time-delays by using angular distributions, we distinguish between the different  angular contributions to the photoelectron wavepacket by selecting the single orbital angular momentum $\ell$ of the many-body wavefunction in the outer region. Thus, the different channels are labeled by the angular momentum of the photoelectron wavepacket $\ell$ uniquely, as we can see in Fig.~\ref{fig:fig5}(c)-(d), and we can isolate them to obtain $\expected{r(t)}$ corresponding to a given $\ell$. Taking all this into account, to obtain the time-delay between the photoelectrons ejected from $2s$ and $2p$ we follow the steps: i) calculate $\expected{r(t)}$ for two different times $t$ and  $\ell=1$ and $2$; ii) obtain $k$ for each channel as described just above Eq.~\eqref{eq:time_delay}; iii) evaluate $t_\text{Coul}(t)$ using the expression~\eqref{eq:time_delay_coul}; iv) calculate $\tau_{2s}$  and $\tau_{2p}$ using the expressions~\eqref{eq:time_delay_ews} and~\eqref{eq:time_delay}; and finally, v) obtain $\tau_{2p-2s}=\tau_{2p}-\tau_{2s}$. 
Let us note that a good description not only of the potential induced by the electrons but also an accurate description of the photoelectron spectrum is mandatory  for the accurate application of this approach. For instance, using a two-electron model in a mean-field potential~\cite{Nagele2012} gives a value of $\tau_{2p-2s}=4.3$~as for a central frequency of 107~eV, which is below the expected theoretical value [see Fig.~\ref{fig:fig6}] although the result qualitatively captures that the electron in the $2p$ shell is emitted after the $2s$. In Fig.~\ref{fig:fig6} we show the streaking time-delay $\tau_{2p-2s}$ assuming a 780~nm IR pulse, to account for  $\tau_\text{CLC}$ for the RAS scheme $(M_1,\,M_2)=(5,0)$ and $(5,4)$, together with calculations which use different methods to account for the electronic correlation~\cite{Feist2014a,Moore2011,PhysRevA.86.061402}. In Ref.~\cite{Moore2011} the $R$-matrix method was employed to describe the streaking process, where the inner region is described using configuration interaction and only one electron is allowed in the outer region. The results of this method is in agreement with the results of the TDHF method, \ie, $(M_1,\,M_2)=(5,0)$, from $90~\text{eV}\le \omega\le 100$~eV, except that  the $R$-matrix prediction is larger for $\omega<90$~eV and lower for $\omega>105$~eV. As stressed in Ref.~\cite{Moore2011}, for $\omega\gtrsim 105$~eV there are contributions from pseudoresonances induced by the expansion in the inner region which may alter the result of the calculation. Furthermore, the sensitivity of the time-delay to the ionization energy of each channel may also interfere for low $\omega$'s. We can also conclude that part of the correlation can be described using a single configuration, as in the TDHF method, due to the flexibility in the propagation provided by the time-dependent orbitals. 
This situation is markedly different from the case we considered in Be where ionization of the Be$[(1s^2 2s^2)^1\text{S}^e]$ ground-state into the channel Be$^+[(1s^2 2p)^2\text{P}^o] + e^- (s \text{ or } d)$ can
not be described by TDHF, since, contrary to the case in Eqs.~(\ref{eq:2p_ejected})-(\ref{eq:2s_ejected}), more than a single orbital in the dominant configuration is changing~\cite{Omiste2017_be}.

The TD-RASSCF method with $(M_1,\,M_2)=(5,4)$ provides a smaller $\tau_{2p-2s}$, which is in agreement with the results of Ref.~\cite{Feist2014a}. In that work, the ground-state and scattering states of Ne are built using up to 38 states to describe the atom and the Ne$^+$ ion to extract the phase shift corresponding to the long- and the short-range interaction, $\sigma_\ell(E)+\delta_\ell(E)$, and hence $\tau_\text{EWS}=\frac{\partial}{\partial E}[\sigma_\ell(E)+\delta_\ell(E)]$. Then, as in our case, the total time-delay is obtained by adding $\tau_\text{CLC}$~\cite{Feist2014a}.  In contrast, in Ref.~\cite{PhysRevA.86.061402}, the time-delay $\tau$ is fully extracted by studying the streaking process in a time-independent diagrammatic approach. The resulting time-delay is $\tau=12$~as for $\omega\sim 105$~eV, a bit higher than the TDHF result. Finally, we note that the difference between the $\tau_\text{2p-2s}$ for the two different RAS schemes, decreases as we increase $\omega$ and 
almost vanishes at $\omega=125$~eV, revealing that the correlation becomes less important with increasing $\omega$ for the considered case in Ne.
Let us note that the EWS time-delay between the $2p$ and $2s$ photoelectrons for $105$~eV, $\tau_{\text{EWS}, 2p}-\tau_{\text{EWS}, 2s}=7.1$ and $5.8$~as for the RAS $(M_1=5,\,M_2=0)$ and $(M_1=5,\,M_2=4)$, are in good agreement with the EWS delay $6.4$~as reported in Ref.~\cite{Schultze2010} and obtained using the state-specific approach for $\omega=106$~eV, but are significantly higher than the $4.0$~as calculated using multiconfigurational Hartree-Fock method, also in Ref.~\cite{Schultze2010}.

\section{Conclusion}
\label{sec:conclusions}
We have applied the TD-RASSCF-D method to investigate the role of electron correlation in the ground-state of Ne, as well as in photoionization processes induced by attosecond XUV linearly polarized laser pulses. We have shown that the TD-RASSCF-D method provides accurate results for the ground-state energy, converging to the MCTDHF method as we increase the electronic correlation,~\ie, the number of accessible configurations.
Most importantly, we also found that it is possible to obtain a better ground-state by increasing the number of orbitals while keeping the number of configurations manageable by 
design of the RAS scheme. The decisive role of the number of orbitals compared with the number of configurations, i.e., the importance of 
having access to a few rather highly excited configurations in the SCF expansion of the total wave function, was also found in cold atom physics~\cite{Leveque2016a}. This finding shows the potential of the TD-RASSCF approach for application to systems where the
MCTDHF can not be practically applied due to the dimension of the problem.
 For the photoionization dynamics, we describe the ionization threshold energy of the channels $\text{Ne}^+~[(1s^22s2p^6){}^2\text{S}^e]$ and $[(1s^22s^22p^5){}^2\text{P}^o]$. We also obtained results for the angular distribution of the electron ejected by the XUV pulse. Moreover, we obtained numerical cross sections which are in agreement with the available experimental data. Finally, we calculated the time-delay between the propagated electrons ejected from the $2p$ and $2s$ shells by taking advantage of the angular momentum decomposition of the photoelectrons to measure independently these two channels. 
 For $\omega=105$~eV, we obtain $\tau_{2p-2s}=9.9$~as in agreement with other theory works~\cite{Kheifets2010,Moore2011,Ivanov2011,Nagele2012,PhysRevA.86.061402,Su2014,Wei2016,Feist2014a} and with the very recent interferometric experimental measurements~\cite{Isinger2017}, but in disagreement with the experimental value of $\sim 21$ as for a photon energy of $106$~eV~\cite{Schultze2010}.
For the present study in Ne, the channels considered are both accessible by the change of a single orbital from the dominant ground-state configuration following single-photon absorption. We found that in this case, the TDHF method gives a qualitatively correct estimate of the time-delay in photoionization. 
In systems that are too complicated to be investigated by 
any theory beyond mean-field, a comparison between, e.g., experimental time-delay data and the results from TDHF could then serve to isolate presence or absence of correlations effects.

\begin{acknowledgments}
 This work was supported by the Villum Kann Rasmussen (VKR) Center of Excellence QUSCOPE. J.J.O. was supported by  NSERC Canada (via a grant to Prof. P. Brumer). The numerical results presented in this work were obtained at the Centre for Scientific Computing, Aarhus.
\end{acknowledgments}

\appendix

\section{Conservation of orbital $m$}
\label{sec:conservation_or_orbital_m}

In this appendix, we prove that the magnetic quantum number $m$ of each orbital is conserved if each orbital is labelled with a well-defined $m$ and the wavefunction is a linear combination of configurations $\left\{\ket{\Phi_{\bm{I}}(t)}\right\}$ with the same total magnetic quantum number $M_L$.

The propagation of the single-electron orbitals is determined by the $\mathcal{Q}$- and~$\mathcal{P}$-space equations, Eqs.~\eqref{eq:q_space} and~\eqref{eq:p_space}, respectively. By setting $\eta_j^i(t)$ either to be $0$ or $h_j^i(t)$, the only contribution in $\mathcal{Q}$-space which may mix the magnetic quantum number $m$ in the orbital $\ket{\phi_i(t)}$ is $\sum_n(\bm{\rho}^{-1}(t))^n_i\sum_{jkl}Q(t)\rho_{nk}^{jl}(t)W_l^k(t)|{\phi_j(t)}\rangle$. 
We first prove that $\left(\bm{\rho}^{-1}(t)\right)_i^n$ is non-zero only for $m_n=m_i$. From the definition of the one-body density operator
\begin{eqnarray*}
  \rho_i^n(t)&=&\sum_{\bm{I},\bm{J}}C_{\bm{I}}(t)^* C_{\bm{J}}(t)\melement{\Phi_{\bm{I}}(t)} {E_i^n} {\Phi_{\bm{J}}(t)},
\end{eqnarray*}
we see that the contribution $\melement{\Phi_{\bm{I}}(t)}{E_i^n}{\Phi_{\bm{J}}(t)}\ne 0$ only if $\ket{\Phi_{\bm{I}}(t)}~\text{and}~\ket{\Phi_{\bm{J}}(t)}$ only differ in the orbitals $\ket{\phi_n(t)}$ and $\ket{\phi_i(t)}$, respectively. Since $M_L$ is the same for all the configurations, ${\rho}_i^n(t)\ne 0\Rightarrow M_L-m_i=M_L-m_n\Rightarrow m_i=m_n$. Thus, ${\rho}_i^n(t)$ is block diagonal in the single-electron magnetic quantum number. This implies that $\bm \rho^{-1}(t)$ is also block diagonal in $m$, and, therefore, $\left(\bm \rho^{-1}(t)\right)^n_i\ne 0$ only if $m_n=m_i$. A similar argument can be applied for $\rho^{jl}_{nk}(t)$ which reads as 
\begin{eqnarray*}
  \rho_{nk}^{jl}(t)&=&\sum_{\bm{I},\bm{J}}C_{\bm{I}}(t)^* C_{\bm{J}}(t)\melement{\Phi_{\bm{I}}(t)} {E_{nk}^{jl}} {\Phi_{\bm{J}}(t)}
\end{eqnarray*}
and it is non-zero only if $m_j+m_l=m_n+m_k$. Finally, since $W_l^k(t)\ket{\phi_j(t)}$ is a function with $m=m_l+m_j-m_k$ and the projector $Q(t)$ preserves the magnetic quantum number, $Q(t)\rho_{nk}^{jl}W_l^k(t)|{\phi_j(t)}\rangle\ne 0$ only if $m_n=m_l+m_j-m_k=m_i$, which ensures the conservation of $m$ by the $\mathcal{Q}$-space equation.

Now, we  prove that the $\mathcal{P}$-space equation~\eqref{eq:p_space} satisfies $\eta_{l'}^{k''}=0$ if $m_{l'}\ne m_{k''}$. First, we evaluate the last summation which involves the two-body elements. It is easy to prove that $v_{kl}^{j''m}(t)\ne 0 \Rightarrow m_k+m_l=m_m+m_{j''}$, and using similar arguments of the previous proof, $\rho_{i'm}^{kl}(t) \ne 0\Rightarrow m_l+m_k=m_{i'}+m_m$. Equating these two expressions, we obtain that $m_{j''}=m_{i'}$, which also holds for the second term of the summation. On the other hand, $A_{k''i'}^{l'j''}(t)= \langle\Psi(t)|E_{i'}^{j''}E_{k''}^{l'}-E_{k''}^{l'}E_{i'}^{j''}|\Psi(t)\rangle=\delta_{k''}^{j''}\rho_{i'}^{l'}-\delta_{i'}^{l'}\rho_{k''}^{j''}\ne 0 \Rightarrow m_{j''}=m_{k''}$ with $i'=l'$ and/or $m_{i'}=m_{l'}$ with $j''=k''$. Taking all this into account when solving this equation, we get that $h_{l'}^{k''}(t)-i\eta_{l'}^{k''}(t)\ne 0$ only if $m_{l'}=m_{k''}$. Finally, using that $h_{l'}^{k''}(t)$ fulfils $m_{l'}=m_{k''}$ it follows that  $\eta_{l'}^{k''}\ne 0\Rightarrow m_{l'}=m_{k''}$.


%

\end{document}